\documentclass[onecolumn,floatfix,superscriptaddress,a4paper,
               showpacs,showkeys,nofootinbib,preprint]{revtex4}
\usepackage{epsfig}
\usepackage{latexsym}
\usepackage{xspace}
\usepackage{hyperref}
\usepackage[latin2]{inputenc}
\usepackage{indentfirst}
\usepackage{enumerate}

\usepackage{amsmath}
\usepackage{amssymb}
\usepackage[english]{babel}
\usepackage{url}

\newcommand{\eq}[1]{\begin{align} #1 \end{align}}

\begin{document}

\title{Statistical Model of the Early Stage of nucleus-nucleus collisions \\
       with exact strangeness conservation
\vspace{0.5cm}}

\author{R.~V. Poberezhnyuk}
\affiliation{Bogolyubov Institute for Theoretical Physics, Kiev, Ukraine}
\author{M. Gazdzicki}
 \affiliation{Goethe--University, Frankfurt, Germany}
 \affiliation{Jan Kochanowski University, Kielce, Poland}
\author{M. I. Gorenstein}
 \affiliation{Bogolyubov Institute for Theoretical Physics, Kiev, Ukraine}
 \affiliation{Frankfurt Institute for Advanced Studies, Frankfurt, Germany
\vspace{0.5cm}}

\begin{abstract}
The Statistical Model of the Early Stage, SMES,
describes a transition between confined and deconfined phases
of strongly interacting matter created in nucleus-nucleus collisions.
The model was formulated in the late 1990s for central Pb+Pb collisions
at the CERN SPS energies. It predicted several signals of the transition
(onset of deconfinement) which were later observed by the NA49 experiment.
The grand canonical ensemble was used to calculate entropy and strangeness
production. This approximation is valid for reactions with mean multiplicities
of particles carrying conserved charges being significantly larger than one.

Recent results of NA61/SHINE on hadron production in inelastic p+p
interactions suggest that the deconfinement may also take place  in
these reactions. However, in this case mean multiplicity of particles
with non-zero strange charge is smaller than one.
Thus for the modelling of p+p interactions
the exact strangeness conservation has to be implemented in the SMES.
This extension of the SMES is presented in the paper.
\end{abstract}

\pacs{12.40.-y, 12.40.Ee}

\keywords{Onset of deconfinement, nucleus-nucleus collisions, proton-proton collisions}

\maketitle

\section{Introduction}
\label{sec-intr}

Strongly interacting matter at sufficiently high energy density is
predicted to exist in a phase of quasi-free quarks and gluons, 
the  quark gluon
plasma (QGP). Relativistic nucleus-nucleus (A+A) collisions provide a unique
opportunity to check this prediction 
and study  properties of the transition to the QGP as well as 
the QGP itself. This is  because the system created in A+A collisions
is close to (at least local) equilibrium. The conclusion is based on 
the success of  statistical
and hydrodynamical models of particle production at high energies
(see e.g. Ref.~\cite{Fl}).
Consequently,  properties of the system (matter) can be characterized
by its equation of state which should include different phases and
transitions between them.
It is important to note that nowadays there is no dynamical understanding
of the observed equilibrium properties of particle production
in A+A collisions.

With increasing collision energy the energy density of matter created at the
early stage of A+A collisions increases.  
Thus, at a sufficiently high collision energy 
the matter is expected  to be created in the QGP phase.
The beginning of the QGP creation with increasing collision energy
is referred to as the onset of deconfinement. 
The experimental search for the onset of deconfinement
in central Pb+Pb collisions was
performed by the NA49 experiment at the Super Proton
Synchrotron (SPS) of the European Organization for Nuclear
Research (CERN) about 15 years ago.  The study was 
motivated~\cite{Afanasev:2000dv} by predictions
of the Statistical Model of the
Early Stage (SMES)~\cite{GG} of A+A collisions.
According to the model the onset of deconfinement in central A+A collisions
should lead to rapid
changes of the energy dependence of several hadron production properties,
all located in a common energy domain.
In particular,
a non-monotonic dependence of the strangeness to entropy ratio as
a function of the collision energy (the {\it horn}) was predicted~\cite{GG}
as an important signal of the transition.
This and other predictions of the SMES were confirmed by  NA49~\cite{:2007fe,:2007fe1}.
Moreover, following results from  the Relativistic
Heavy Ion Collider  at Brookhaven National Laboratory
and the Large Hadron Collider (LHC) at CERN agree with
the NA49 results and their interpretation (see Ref.~\cite{rustamov}).
The SMES  predictions and the experimental evidence for
the onset of deconfinement are presented in
recent reviews~\cite{review}.

The SMES  is probably the 
simplest model of the onset of deconfinement. 
This leads to a number of advantages and disadvantages.
In particular, the SMES is frequently criticized
for being based on simple assumptions which can not 
be justified within popular dynamical approaches 
to A+A collisions. 

In this paper we concentrate on a single aspect of the SMES
which concerns the finite size effects for
strange hadron production. The SMES predictions for  strangeness production were
calculated within the grand canonical ensemble (GCE).
This approximation is  valid for central Pb+Pb  collisions at the
SPS energies in which mean multiplicity of particles with non-zero
strange charge is significantly larger than one.
However, this is not the case for
inelastic p+p interactions
at the SPS energies. Here the exact strangeness conservation
has to be imposed using the canonical  ensemble (CE)~\cite{CE,CE1,CE2,CE3}.

\begin{figure}[tbh]
\centering
\includegraphics[width=0.49\columnwidth]{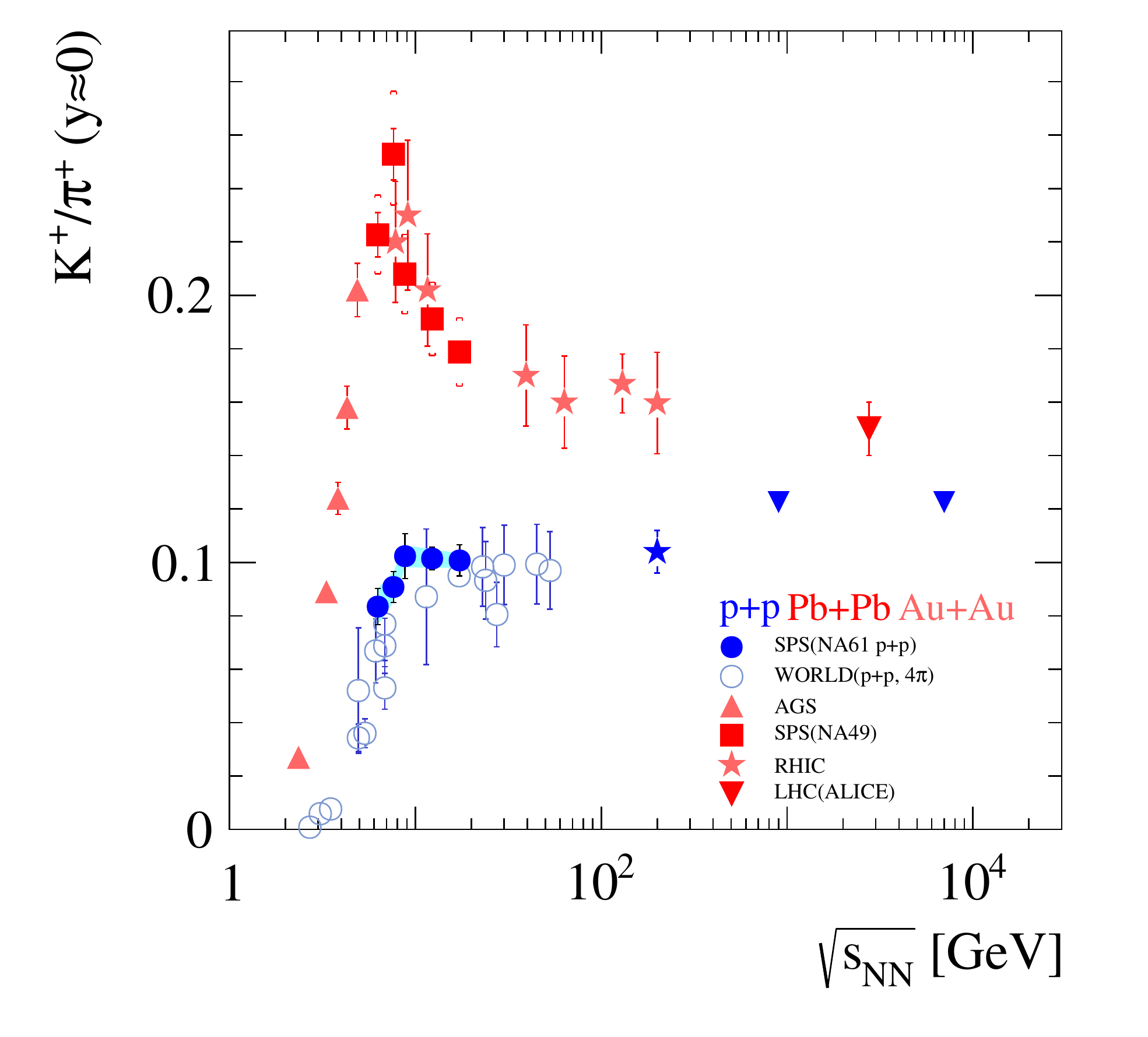}
  \caption{\label{fig:onset}
The {\it horn}  structure
in the energy dependence of the  K$^+/\pi^+$ ratio
is interpreted as evidence for
the onset of deconfinement located at low CERN SPS energies.
The structure was first discovered by NA49 in central Pb+Pb collisions.
Surprisingly its shadow is visible in inelastic p+p interactions as indicated
by the new NA61/SHINE data.
}
\end{figure}

Recently, the NA61/SHINE Collaboration at the CERN SPS published results
on hadron production in p+p interactions~\cite{NA61, NA61-s}.
They suggest that in these reactions the strangeness to entropy ratio
(experimentally replaced by the $K^+$ to $\pi^+$ ratio)
also changes rapidly in the SPS energy range, see Fig.~\ref{fig:onset}.
However, the ratio and its energy dependence are significantly
different from the {\it horn} measured in central Pb+Pb collisions.
Can these results be explained by the onset of deconfinement as modelled by the SMES?
The first step towards an answer to this question is taken in this paper
by introducing the exact strangeness conservation.
In order to allow for a direct comparison with the previously published
predictions, the remaining SMES assumptions, parameters and notations are
kept unchanged.

The paper is organized as follows.  In Sec.~\ref{sec-GCE} the GCE
formulation of the SMES is briefly recapitulated.
The exact strangeness conservation is introduced in the SMES in
Sec.~\ref{sec-CE} and results for p+p interactions and collisions of small nuclei
are presented.
A summary in Sec.~\ref{sec-sum} closes the article.

\section{The SMES model in brief}
\label{sec-GCE}

The SMES model was formulated almost 20 years ago. Its basic assumptions,
parameters and results are summarized in this section.
Together with the notation used in the original paper
they are here kept unchanged as much as possible in order to allow for
a direct comparison with the previously published results.

The SMES assumes that the matter created at the early stage of collisions
has zero conserved charges.  Consequently, its properties
are entirely defined by the available energy and the volume in which
production takes place. In central A+A collisions this volume is chosen as the Lorentz
contracted volume occupied by the colliding nucleons (participant nucleons) from a single
nucleus:
\eq{\label{volume}
V=\frac{4 \pi r^3_0 A_p/3}{\sqrt{s_{NN}}/2m_N}~,~~
}
where $m_N$ is the nucleon mass, $\sqrt{s_{NN}}$ is the center of mass energy of the nucleon pair,
$A_p$ is the number of participant nucleons
from a single nucleus. The $r_0$ parameter is taken to be 1.30~fm in order to fit the mean
baryon density in the nucleus, $\rho_0 = 0.11$~fm$^{-3}$.

Only a fraction, $\eta$, of the total energy in A+A collisions is transformed into the energy
of new degrees of freedom created at the early stage. This is because a part of the energy
is carried by the net baryon number. The released
(inelastic) energy  is expressed as
\eq{\label{E}
E~=~\eta \,(\sqrt{s_{NN}}~-~2m_N)\,A_p~,
}
where
the parameter $\eta$ is assumed to be independent of
the collision energy and the system size.
The value of $\eta$ used for  numerical calculations is 0.67~\cite{GG}.

Assumptions~(\ref{volume}) and~(\ref{E}) with $\eta = 1$ correspond
to the Landau hydrodynamical model~\cite{landau}.
Similarly to this model, in the SMES we do not consider 
dynamical mechanisms leading
to a fast thermalization of the  matter.
The SMES model postulates that the creation of new
particles at the early stage of collision is a statistical process,
namely, all microscopic states allowed by
conservation laws is equally probable.

The SMES predictions for the pion multiplicities are based on 
the assumption that the entropy
generated at the early stage of collision is (approximately) 
conserved during the expansion of
produced matter. It was indeed observed that the dissipative effects estimated by the
ratio of the shear viscosity to the entropy density 
are small for the strongly interacting matter, 
especially in a region of the deconfinement transition
(see, e.g.,~\cite{eta} and references therein). It should be also noted that particle
interactions play rather different role for the equilibrium properties (e.g., the equation of state)
and the kinetic coefficients (e.g., the shear viscosity). This is clearly demonstrated
by a simple example of the hard balls system~\cite{eta1}. 
The hard core particle radius $r$ leads 
to small corrections to the ideal gas equation of state due to the excluded
volume effects, but the shear viscosity
as it behaves as $\propto r^{-2}$ and thus it is strongly dependent on~$r$.

The elementary particles of strong interactions are quarks and gluons. The deconfined
state is considered to be composed of $u$, $d$ and $s$ quarks and the corresponding
anti-quarks each with internal number of degrees of freedom equal to 6 (3 color states and 2
spin states). The contribution of $c$, $b$ and $t$ quarks is neglected due to their large
masses. The internal number of degrees of freedom for gluons is 16 (8 color states and 2
spin states). The masses of gluons and
non-strange (anti)quarks are taken to be 0.  The strange (anti)quark mass is taken to be
175~MeV~\cite{GG}. The properties of equilibrated
matter are characterized by an equation of state (EoS). For the case of quarks and
gluons the bag model EoS is used~\cite{bag}, i.e., the ideal gas EoS modified by a
bag constant $B$.  This equilibrium state of quarks and gluons is called the Quark Gluon
Plasma or Q state.

 The SMES uses an effective parametrization of the confined hadron state, denoted as
W state. The non-strange degrees of freedom which dominate the entropy
production are taken to be massless bosons. Their internal number of degrees
of freedom is taken to be 16 i.e., about 3 times lower than
the internal number of effective degrees of freedom in the QGP.  The
mass of strange degrees of freedom is assumed to be 500~MeV, equal to the kaon mass.
The internal number of strange degrees of freedom is assumed to be 14.
For the W-state the ideal gas
EoS is selected. Clearly, this description of the confined state should only be treated as an
effective parametrization. The numerical values of the parameters are fixed by fitting A+A data
at the AGS, see for details Ref.~\cite{GG}.

The model assumes that always the maximum entropy state is created at the
early stage of A+A collisions. In the model with two different states (W and Q)
the form of maximum state changes with the collision energy. The regions in which
the equilibrium state is in the form of a pure W or a pure Q state, are
separated by the region in which both states coexist (the mixed phase).
The maximum entropy condition is equivalent to the
assumption of the first order phase transitions with
the Gibbs criterion for the mixed phase (see Appendix B in Ref.~\cite{GG}).
Namely
at a given temperature $T$ the system occupies a pure phase W or Q
whose pressure is larger,
the mixed phase is formed if both pressures are equal $p_W=p_Q$.
The  transition temperature between the W  and Q phases
is assumed to be $T_c=200$~MeV.

Using the assumptions and parameters defined above
predictions of the SMES can be calculated.
The early stage energy density reads:
\eq{\label{epsilon}
\varepsilon~\equiv~\frac{E}{V}=\frac{\eta \rho_0 (\sqrt{s_{NN}}-2 m_N) \sqrt{s_{NN}}}{2 m_N}~.
}

The pressure and energy density functions in the W-phase and Q-phase are equal to:
\eq{\label{p-W}
&
p_W(T)~=~\frac{\pi^2 g_W}{90}T^4~+~
\frac{g^s_W}{2 \pi^2} \int^{\infty}_0 \frac{dk\,k^4}{3(k^2+m_W^2)^{1/2}}\exp\left[{-~\frac{(k^2+m^2_W)^{1/2}}{T}}\right]~,\\
& \varepsilon_W(T)~=~\frac{\pi^2 g_W}{30}T^4+\frac{g^s_W}{2 \pi^2} \int^{\infty}_0 dk\, k^2 (k^2+m_W^2)^{1/2}\exp\left[{-~\frac{(k^2+m^2_W)^{1/2}}{T}}\right]~,\label{e-W}\\
&p_Q(T)~=~\frac{\pi^2 g_Q}{90}T_c^4~+~
\frac{g^s_Q}{2 \pi^2} \int^{\infty}_0 \frac{dk~k^4}{3(k^2+m_Q^2)^{1/2}}\exp\left[{-~\frac{(k^2+m^2_Q)^{1/2}}{T}}\right]~
-~B~,
\label{p-Q}\\
&\varepsilon_Q(T)~=~\frac{\pi^2 g_Q}{30}T^4~+~
\frac{g^s_Q}{2 \pi^2} \int^{\infty}_0 dk\, k^2 (k^2+m_Q^2)^{1/2}\exp\left[{-~\frac{(k^2+m^2_Q)^{1/2}}{T}}\right]
~+~B~.\label{e-Q}
}
The strange particle contribution to
thermodynamical functions (\ref{p-W}-\ref{e-Q})
are taken within the Boltzmann approximation. This simplification is important
for the CE treatment which will be discussed in the next section. Note that
in Ref.~\cite{GG} the Fermi distribution with $m^*_Q=175$~MeV was used for the strange
quarks. In order to minimize differences to the previous results we choose here a larger value
of $m_Q = 216.5$~MeV which  leads to the same number of strange quarks at the phase
transition temperature ($T_c=200$~MeV):
\eq{\label{Fermi}
\int_0^{\infty} k^2dk\,\exp \left(-\,\sqrt{k^2+m_Q^2}/T_c\right)~=~
\int_0^{\infty} k^2dk\,\left[\exp \left(\sqrt{k^2+m_Q^{*2}}/T_c\right)~+~1\right]^{-1}~.
}
%
%
Then the bag constant  $B=570$~MeV/fm$^{3}$ is calculated using
the Gibbs criterion of equal pressures:
\eq{\label{G}
p_W(T_c)~=~p_Q(T_c)~.
}

The entropy densities in the pure phases ($i=$W, Q) read:
\eq{\label{s}
s_i(T)~=~\frac{p_i(T)~+~\varepsilon_i (T)}{T}~.
}
%
%
In the mixed phase
the W and Q phases coexist. The fraction of volume occupied by the Q phase is denoted as $\xi$.
The energy and entropy densities in the mixed phase are
\eq{\label{e-mix}
&\varepsilon_{\rm mix}(T_c)~=~\xi\,\varepsilon_Q (T_c)~+~(1-\xi)\,\varepsilon_W (T_c)~,\\
%
%
&s_{\rm mix}(T_c)~=~\xi\,s_Q (T_c)~+~(1-\xi)\,s_W (T_c)~.\label{s-mix}
}
%
\begin{figure}[!htb]
\includegraphics[width=0.49\textwidth]{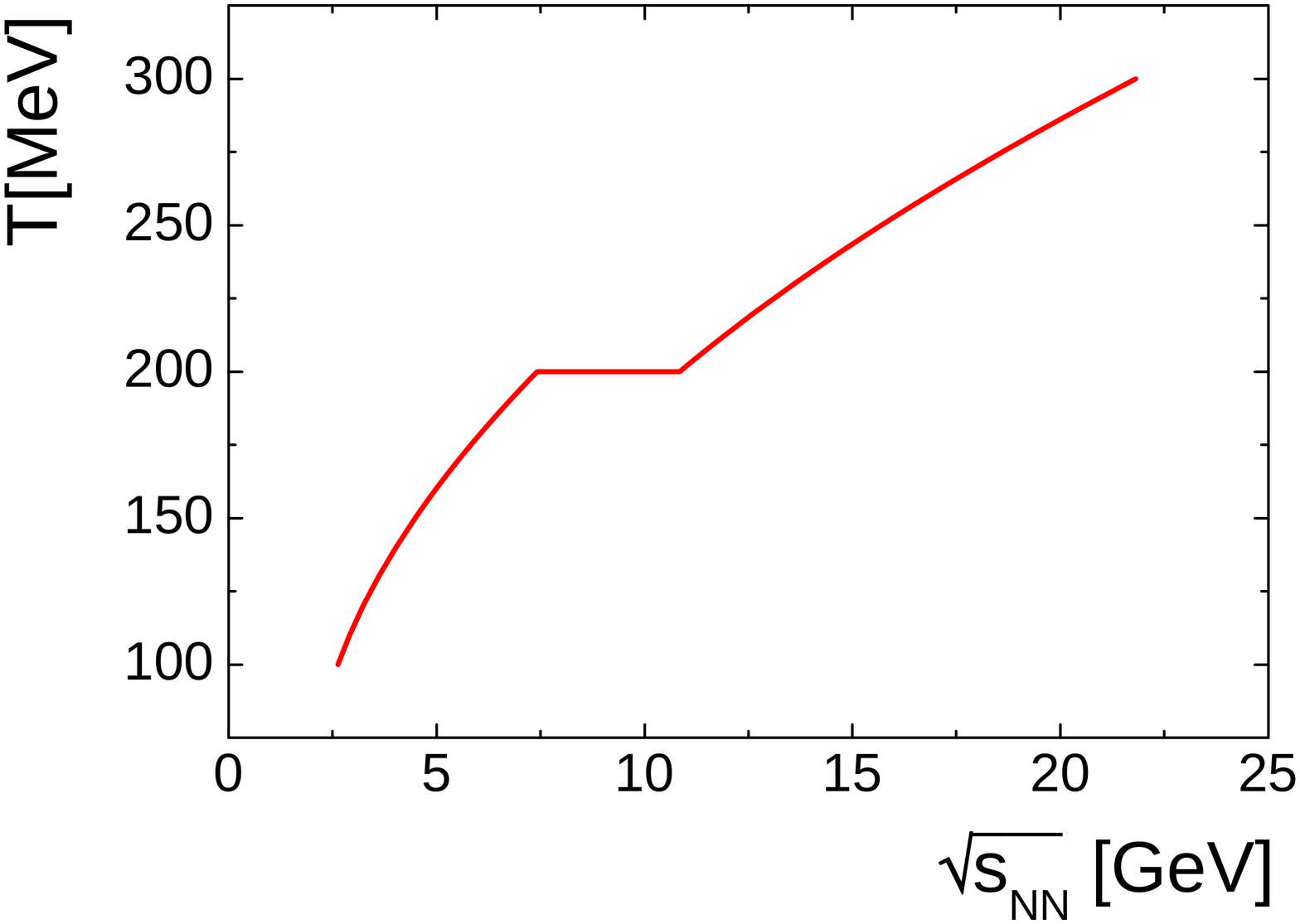}
\includegraphics[width=0.49\textwidth]{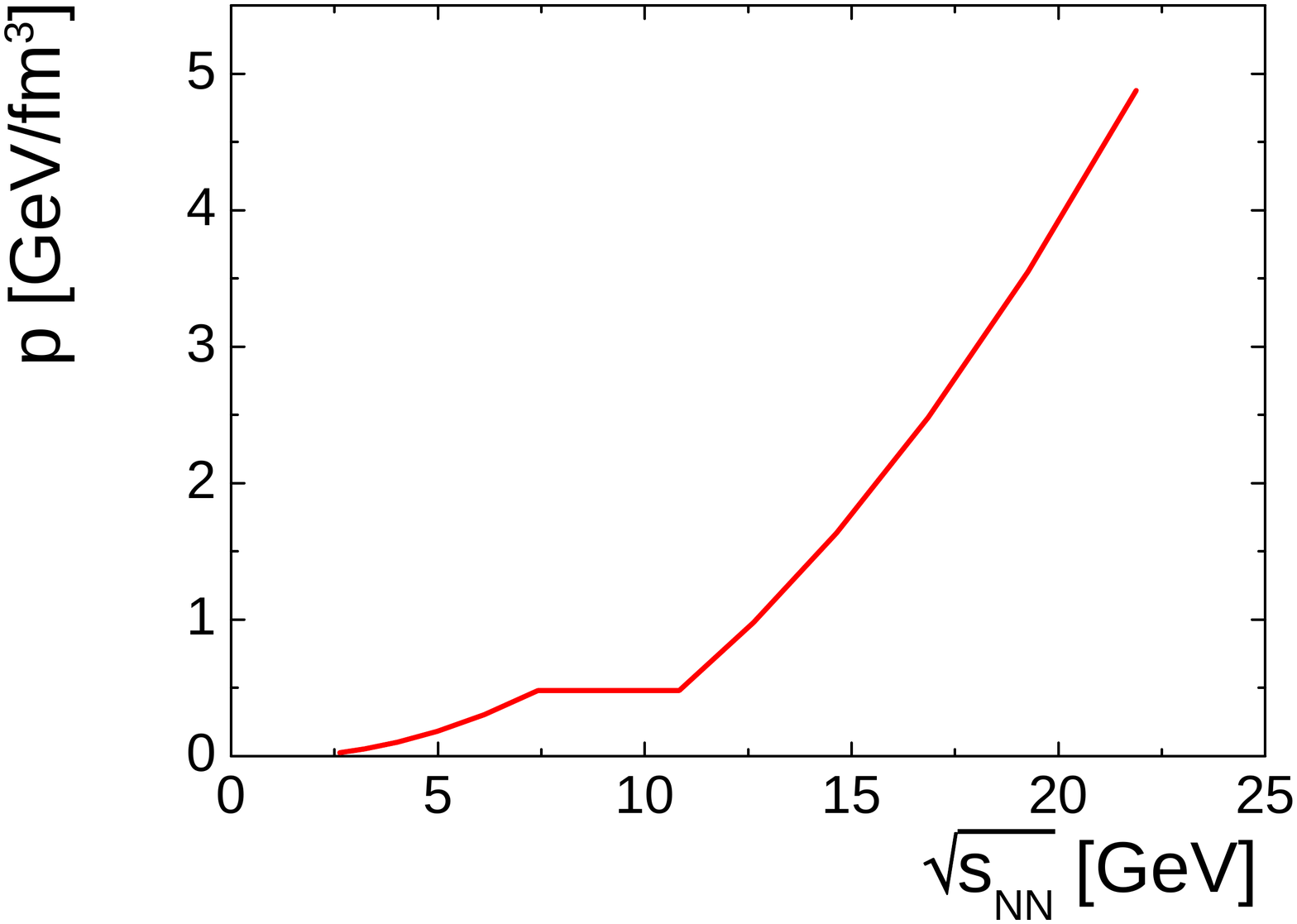}
\caption{The temperature ($left$) and pressure ($right$) of the matter
created at the early stage of A+A collisions as  function of  collision energy. 
\label{fig-T-GCE}
}
\end{figure}

%
%
\begin{figure}[!htb]
\includegraphics[width=0.49\textwidth]{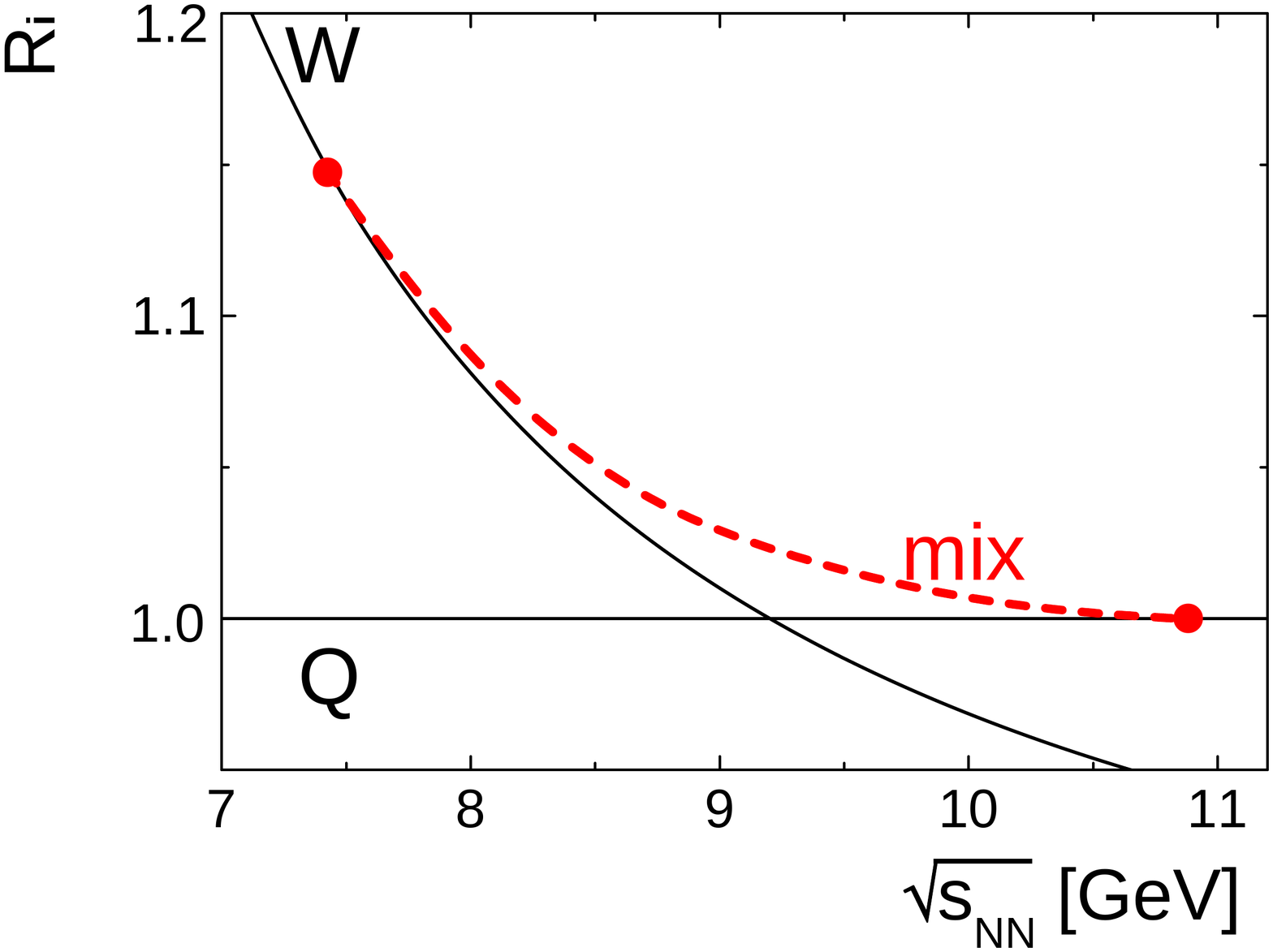}
\includegraphics[width=0.49\textwidth]{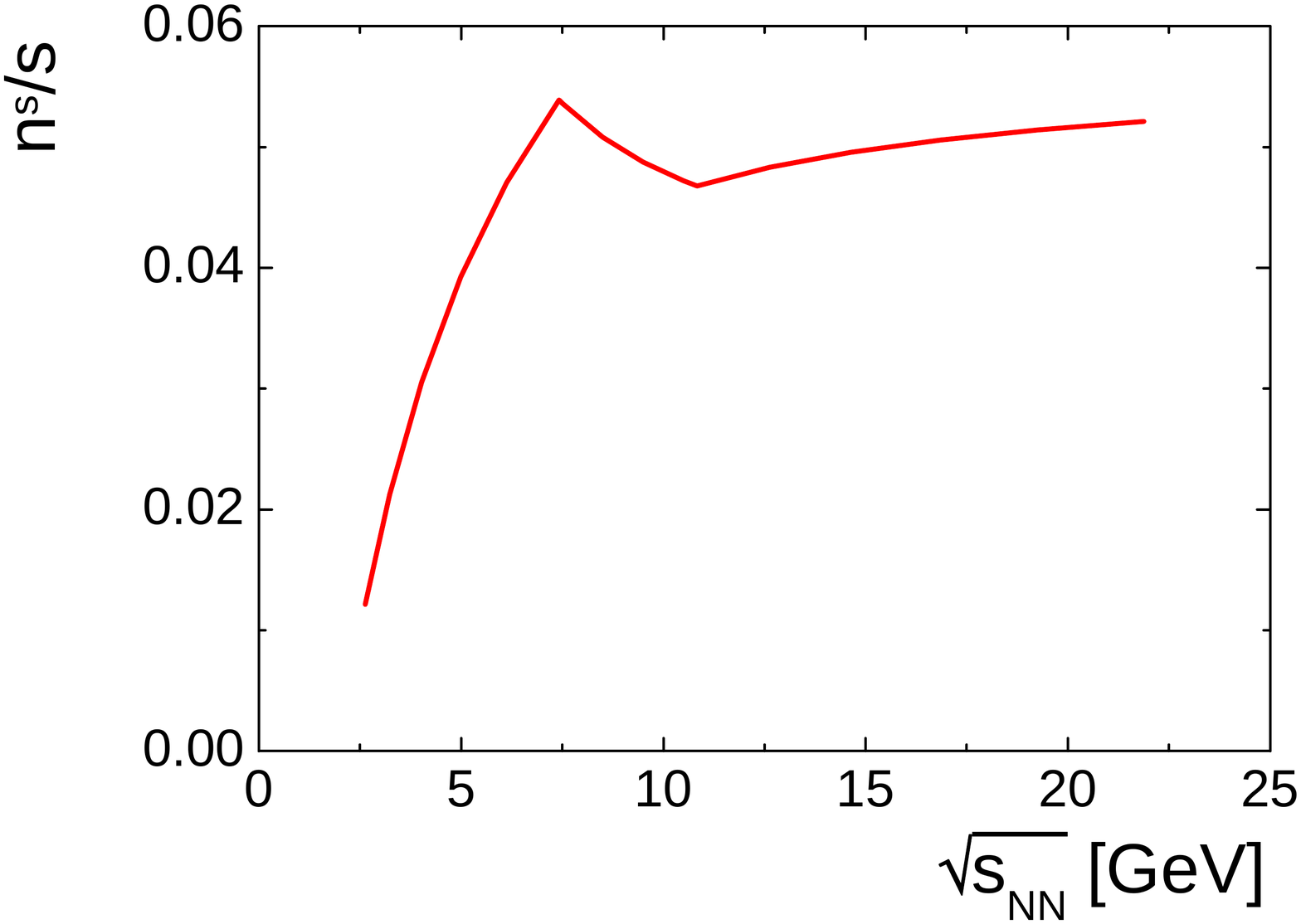}
\caption{$Left$: The ratio of entropy densities $s_i/s_Q$ with $i$ referring
to  the W (solid line), Q (horizontal solid line), and mixed (dashed line) phases,
as a function of  collision energy. The full circles
correspond to the beginning and end of the mixed phase
given by Eq.~(\ref{s1s2}). $Right$: The strangeness to entropy
ratio $n^s/s$ as a function of  collision energy.
\label{fig-smix-GCE}
}
\end{figure}
The temperature $T$ and pressure $p$ are shown
as a functions of the collision energy in Figs.~\ref{fig-T-GCE} $left$ 
and $right$, respectively.
The mixed phase starts at collision energy $\sqrt{s_{NN,1}}$ and ends at $\sqrt{s_{NN,2}}$:
\eq{\label{s1s2}
\sqrt{s_{NN,1}}~=~7.42~{\rm GeV},~~~~\sqrt{s_{NN,2}}~=~10.83~{\rm GeV}~.
}
The equivalence of the  Gibbs criterion and
the maximum entropy condition
is illustrated in Fig.~\ref{fig-smix-GCE} $left$, where the ratios
$R_i=s_i/s_Q$ are presented for $i=$W, mix, and Q.

%

The number density of the sum of strange and anti-strange particles in the GCE
can be calculated as

\eq{\label{nW-s}
& n^s_W(T)~=~\frac{g^s_W}{2 \pi^2} \int^{\infty}_0 dk\,k^2 \exp\left[-~\frac{(k^2+m^2_W)^{1/2}}{T}\right] \\
& n^s_Q(T)~=~ \frac{g^s_Q}{2 \pi^2} \int^{\infty}_0 dk\,
k^2 \exp\left[-~\frac{(k^2+m^2_Q)^{1/2}}{T}\right]~,\label{nQ-s}\\
& n^s_{mix}(\xi)~=~\xi\,n^s_Q (T_c)~+~(1-\xi)\,n^s_W (T_c)~.\label{n-mix}
}
In Fig.~\ref{fig-smix-GCE} $right$ the strangeness to entropy ratio,
$n_s/s$, is shown as a function of the collision energy.

\section{phase transition with exact strangeness conservation}
\label{sec-CE}

In p+p interactions at the CERN SPS energies mean multiplicity of
produced strange and anti-strange particles is smaller than one.
Thus in this case the exact strangeness conservation  should be
taken into account.
In the statistical models this is done within the CE formulation.
The CE partition function of strange particles assures
an equal number of strange and
anti-strange charges, $N_s=N_{\overline{s}}$, in each microscopic state
of the system. For the W and Q phases it has a similar form and reads
\eq{\label{Z-CE}
Z_{ce}(T,V,\lambda)=
\sum_{N_s=0}^{\infty}\sum_{N_{\overline{s}}=0}^\infty
\frac{z^{N_{s}}}{N_s!}\,
\frac{z^{N_{\overline{s}}}}{N_{\overline{s}}!}\,
\delta(N_s-N_{\overline{s}})
=\frac{1}{2\pi}\int_0^{2\pi}d\phi\,
\exp\Big[z\left(e^{i\phi}+ e^{-i\phi}\right)\Big]=I_0(2z)~,
}
where
\eq{\label{z}
z~=~z_{W,Q}~=~\lambda~\frac{1}{2}\,V\,n^s_{W,Q}(T)~.
}
The auxiliary $\lambda$ parameter in Eq.~(\ref{z})
is introduced
to calculate the total strangeness density in the CE:
\eq{\label{ns-CE}
n_{W,Q}^{s(CE)}(T,V)~=~\frac{1}{V}\,
\Big[\frac{\partial \ln Z_{ce}}
{\partial \lambda}\Big]_{\lambda=1}~
=~~n^{s}_{W,Q}(T) ~\frac{I_1\left[V n^{s}_{W,Q}(T)\right]}{I_0\left[V n^{s}_{W,Q}(T) \right]}~.
%
%
}
The ratio of Bessel
functions $I_1$ and $I_0$ in Eq.~(\ref{ns-CE}) quantifies the
strangeness suppression (relatively to the GCE yield) due to the conservation of  net
strangeness in each microscopic state of the CE.

In order to take into account the exact strangeness conservation
for  thermodynamical functions
it is convenient to rewrite them as following:
\eq{
& p^{({\rm CE})}_W(T,V)~=~\frac{\pi^2 g_W}{90}T^4~+~T\, n^{s({\rm CE})}_W (T,V)~, \label{pW-CE}\\
& \varepsilon^{({\rm CE})}_W(T,V)~=~\frac{\pi^2 g_W}{30}T^4~+~\omega_W(T,V)\, n^{s({\rm CE})}_W (T,V)~, \label{eW-CE}\\
& p^{({\rm CE})}_Q(T,V)~=~\frac{\pi^2 g_Q}{90}T^4~+~T\, n^{s({\rm CE})}_Q (T,V)~-~B~,\label{pQ-CE}\\
& \varepsilon^{({\rm CE})}_Q(T,V)~ =~\frac{\pi^2 g_Q}{30}T^4~+~\omega_Q(T)\, n^{s({\rm CE})}_Q (T,V)~+~B~, \label{eQ-CE}
}
where $n^{s({\rm CE})}_{W,Q} (T,V)$ is given by Eq.~(\ref{ns-CE}),
and $\omega_{W,Q}(T)$ is average energy of strange particle:
\eq{\label{omega}
\omega_{W,Q}(T)~=~\frac{\int^{\infty}_0 dk\, k^2 (k^2+m_{W,Q}^2)^{1/2}
\, \exp\left[-~(k^2+m_{W,Q}^2)^{1/2}/T\right] }
{ \int^{\infty}_0 dk\, k^2
\exp\left[-~(k^2+m_{W,Q}^2)^{1/2}/T\right] }~,
}
with $m_W$  taken in the W phase, and $m_Q$ in the Q phase.
The entropy density is given by Eq.~(\ref{s}).

For  $A_p\gg 1$ the system volume (\ref{volume}) is large, and $Vn_{W,Q}^s\gg 1$.
Then one finds that
$I_1\left[V n^{s}_{W,Q}\right]/I_0\left[V n^{s}_{W,Q}\right] \rightarrow 1$ and, therefore,
$n^{s({\rm CE})}_{W,Q}\rightarrow n^{s}_{W,Q}$. The results for
the CE and GCE become equivalent in this
thermodynamical limit, and Eqs.~(\ref{pW-CE}-\ref{eQ-CE})  coincide with
Eqs.~(\ref{p-W}-\ref{e-Q}).

In the mixed phase, Eq.~(\ref{ns-CE}) should be
replaced by
\eq{\label{ns-CE-mix}
n^{s,{\rm mix}}_{W,Q}(T,V,\xi)~=~n^{s}_{W,Q}(T) ~\frac{I_1[X]}{I_0[X]}~,
}
where
\eq{\label{X}
X~=~X(T,V,\xi)~ =~\xi\,V\, n^{s}_{Q}(T)~+~(1-\xi)\,V\, n^{s}_{W}(T)~
}
is the total GCE number of strange and anti-strange particles
(both hadrons and quarks) in the mixed phase. This is because
the CE condition of  zero net strangeness in the mixed phase
should be obeyed by the whole system and not by its phases separately.

At each $\sqrt{s_{NN}}$ one calculates $V$ and $\varepsilon$
according to Eqs.~(\ref{volume}) and (\ref{epsilon}), respectively.
The CE Eqs.~(\ref{ns-CE}-\ref{eQ-CE}) are used to obtain results
for the pure W and Q  phases.
In the mixed phase, the temperature
$T$ and the parameter $\xi$ are obtained by solving the equations:
\eq{\label{e-CE-mix}
& \xi\,\varepsilon_Q^{\rm mix}[T,X]~+~(1-\xi)\,\varepsilon_W^{\rm mix}[T,X]~
=~\varepsilon(\sqrt{s_{NN}})~,\\
& p^{\rm mix}_Q[T,X]~=~p_W^{\rm mix}[T,X]~,\label{p-CE-mix}
}
where $\varepsilon_{W,Q}^{\rm mix}$ and $p_{W,Q}^{\rm mix}$ are
given by Eqs.~(\ref{pW-CE}-\ref{eQ-CE}) with $n^{s,{\rm mix}}_{W,Q}$ (\ref{ns-CE-mix})
instead of $n^{s,{\rm (CE)}}_{W,Q}$ (\ref{ns-CE}).

The collision energy $\sqrt{s_{NN,1}}$ and temperature $T_1$,
where the mixed phase starts, and $\sqrt{s_{NN,2}}$ and $T_2$, where the mixed phase
ends, are obtained as solutions of Eqs.~(\ref{e-CE-mix},\ref{p-CE-mix}) for $\xi=0$ and $\xi=1$,
respectively.
One finds:
\eq{\label{mix-CE-1}
& T_1~=~203.4~{\rm MeV}~,~~~~~~\sqrt{s_{NN,1}}~=~7.20~{\rm GeV}~, \\
& T_2~=~202.9~{\rm MeV}~,~~~~~~\sqrt{s_{NN,2}}~=~10.75~{\rm GeV}~.\label{mix-CE-2}
}

\begin{figure}[!htb]
\includegraphics[width=0.49\textwidth]{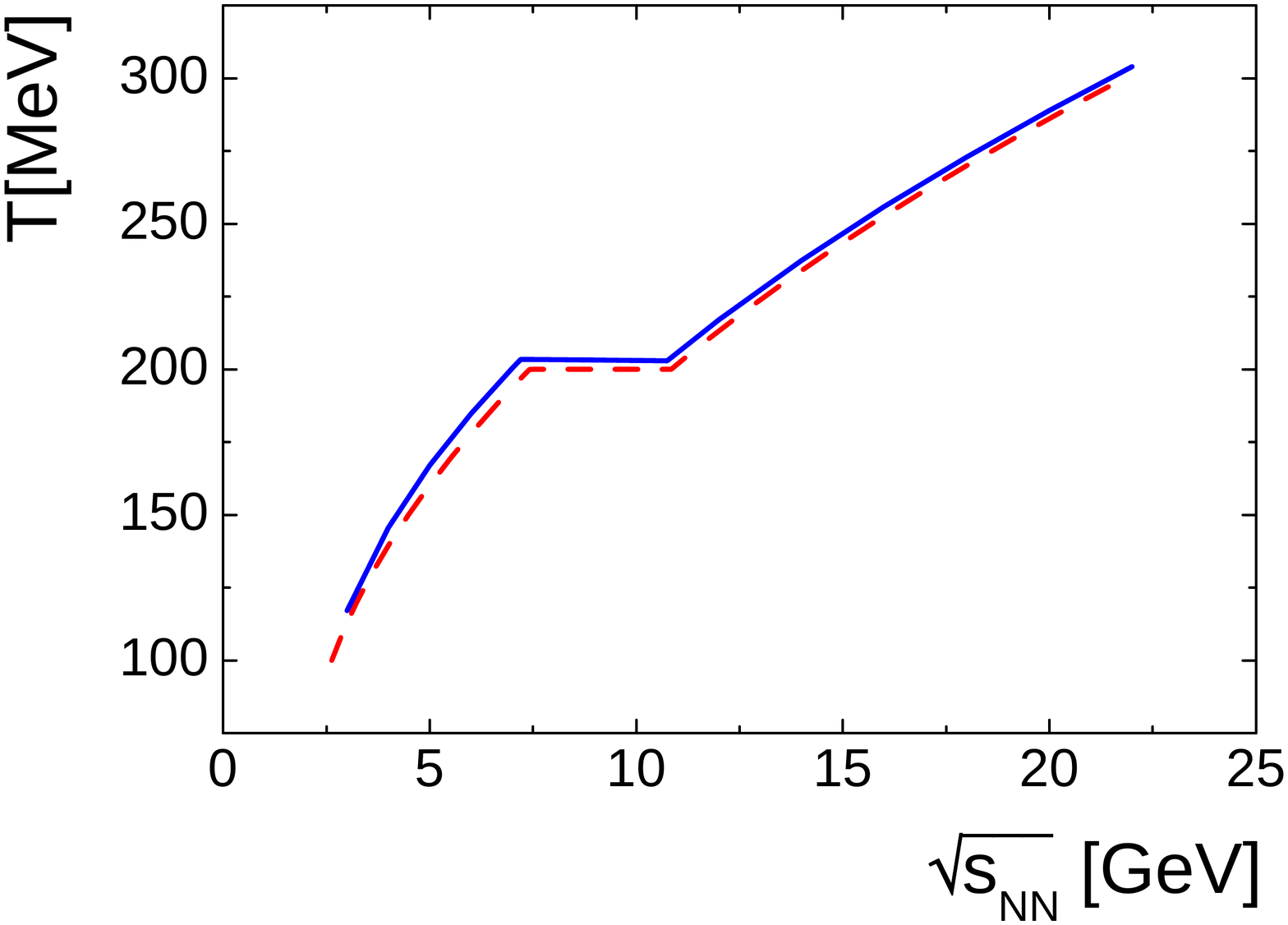}
\includegraphics[width=0.49\textwidth]{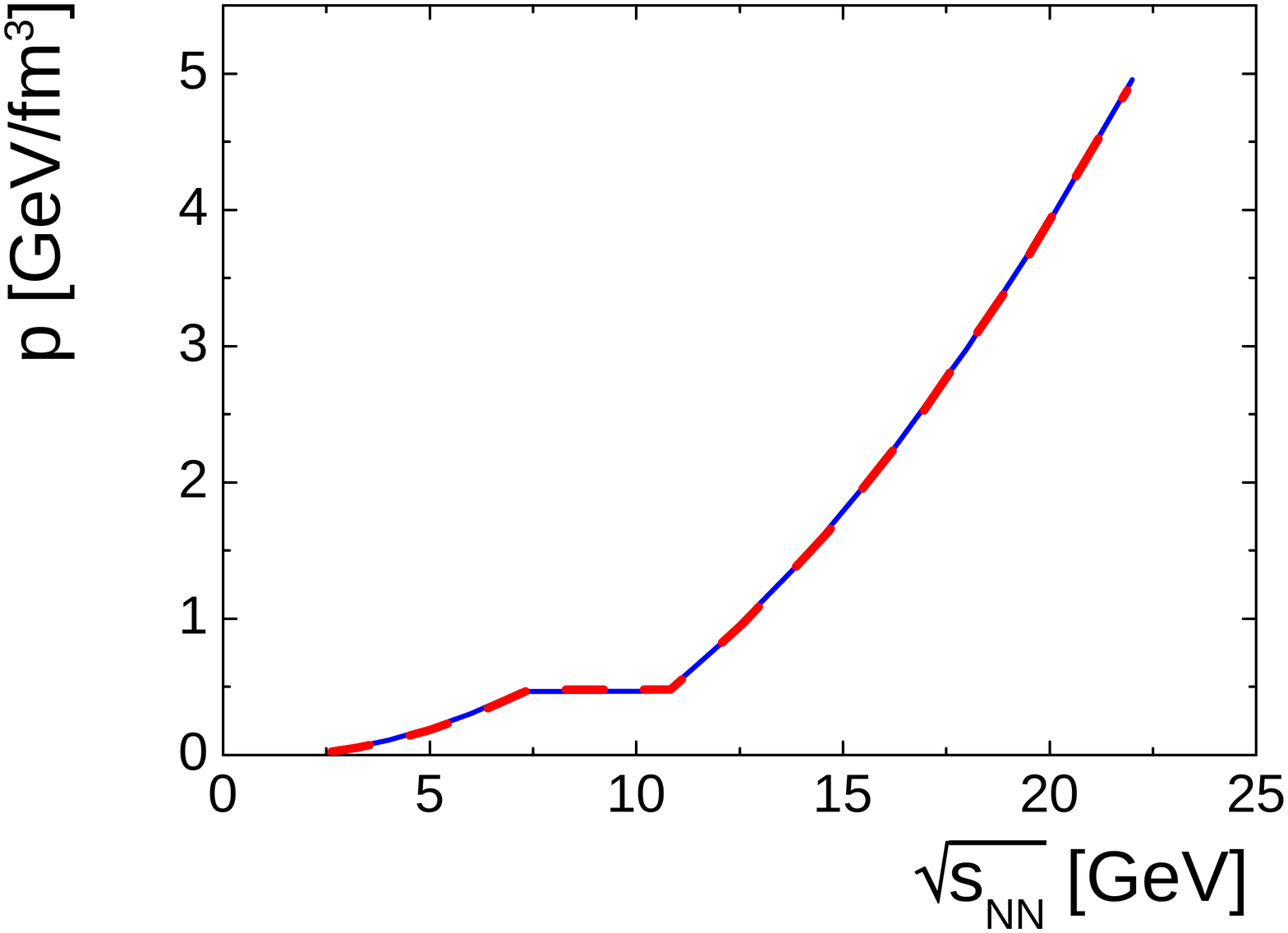}
\caption{The CE temperature ($left$) and pressure ($right$) as  function of 
collision energy
are shown by the solid lines. The dashed lines correspond to the GCE result
presented in Fig.~\ref{fig-T-GCE}.
\label{fig-T-CE}
}
\end{figure}

The collision energy dependence of $T$ and $p$ obtained within the CE
is shown by the solid lines
in Fig.~\ref{fig-T-CE} $left$ and $right$, respectively.
The dashed lines in Fig.~\ref{fig-T-CE}
correspond to the GCE results presented in Fig.~\ref{fig-T-GCE}.
The CE and GCE curves are similar.
A slightly larger value of $T$ in the CE than in the GCE  is needed
to compensate the CE suppression of energy density. Note that $\varepsilon$ as a function
of $\sqrt{s_{NN}}$ is given by Eq.~(\ref{epsilon}) and, thus, it is independent of
the system size.
The entropy density is given by Eq.~(\ref{s})
in terms of $p$, $\varepsilon$, and $T$. Therefore, the entropy density $s$
is weakly affected by the exact strangeness
conservation imposed in the CE.

\begin{figure}[!htb]
\includegraphics[width=0.49\textwidth]{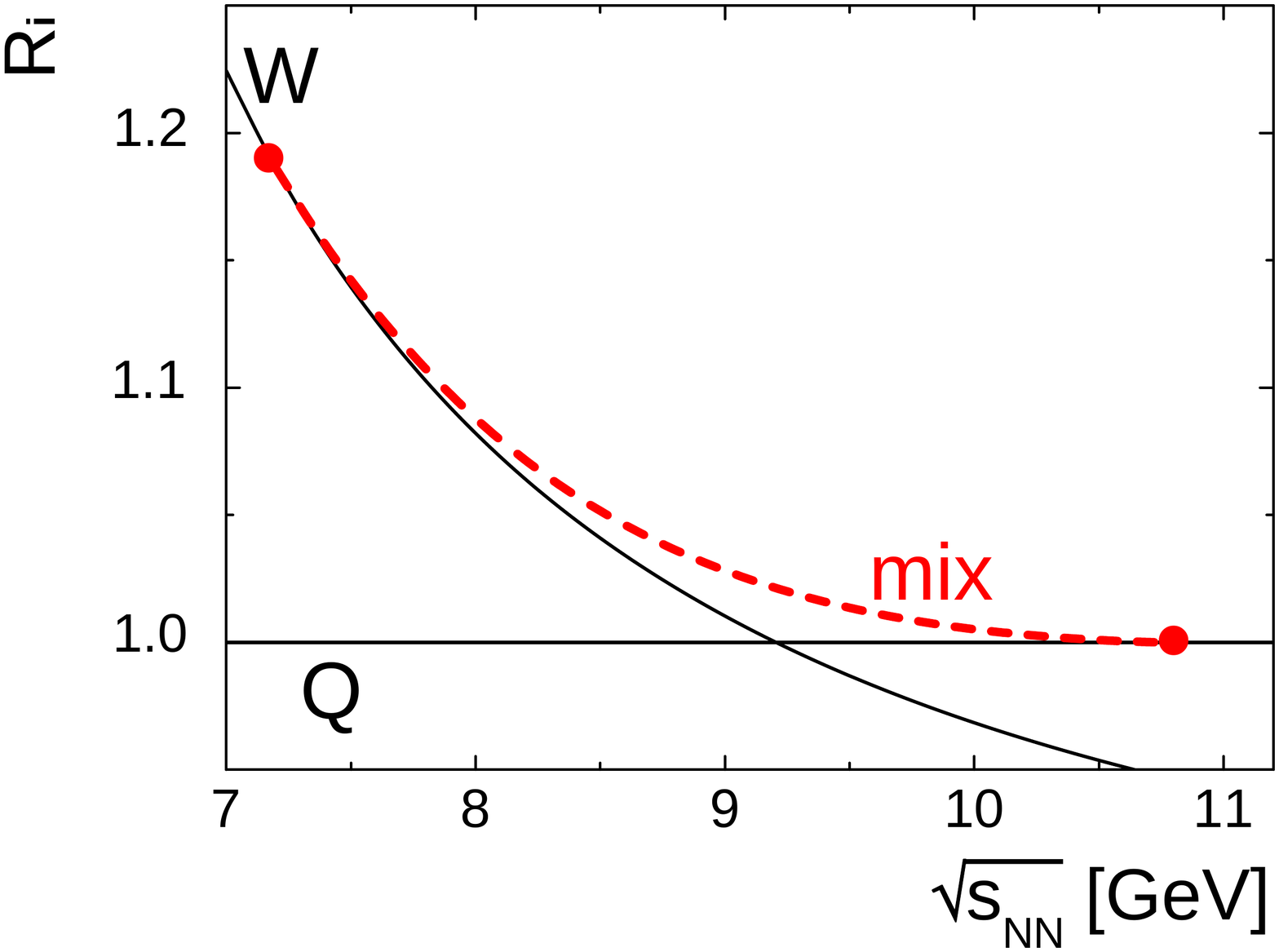}
\includegraphics[width=0.49\textwidth]{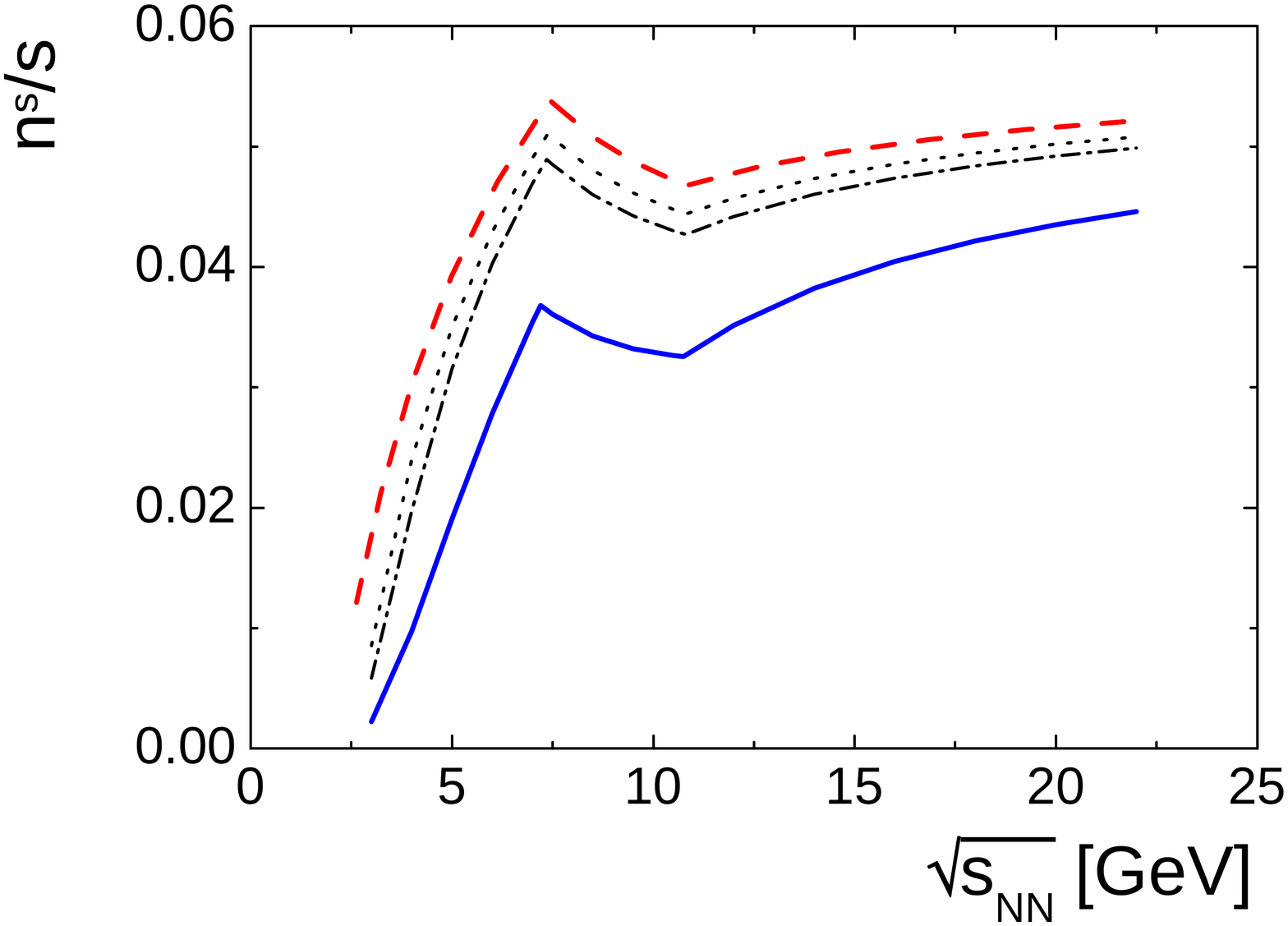}
\caption{ (a): The CE ($A_p = 1$) ratio of entropy densities $s_i/s_Q$ with $i$ referring
to  the W (solid line), Q (horizontal solid line), and mixed (dashed line) phases
as a function of the collision energy.
The full circles correspond to the beginning and end of the mixed phase
given by Eq.~(\ref{mix-CE-1}) and Eq.~(\ref{mix-CE-2}), respectively.
(b): The CE  strangeness to entropy ratio as a function of the collision energy.
The solid line corresponds
to $A_p=1$ and the dashed line to $A_p\gg 1$ which coincides with the GCE
results presented in Fig.~\ref{fig-smix-GCE} (b). The dashed-dotted and dotted
lines show the CE results for $A_p=3$ and 5, respectively.
\label{fig-horn-pp}
}
\end{figure}

The  Gibbs criterion (\ref{p-CE-mix}) used in the CE is again equivalent
to the maximum entropy condition.
This is illustrated in Fig.~\ref{fig-horn-pp} $left$, where the ratios
$R_i=(s_i/s_Q)_{\rm CE}$ calculated in the CE are presented for entropies
$s_H$, $s_Q$ and $s_{mix}$ for $A_p = 1$.

Figure~\ref{fig-horn-pp} $right$ presents  energy dependence of
the strangeness to entropy ratio, $n^s/s$, calculated within the CE
for $A_p =1, 3$ and 5 as well as the result for the GCE ($A_p\gg 1$).

\begin{figure}[!htb]
\includegraphics[width=0.60\textwidth]{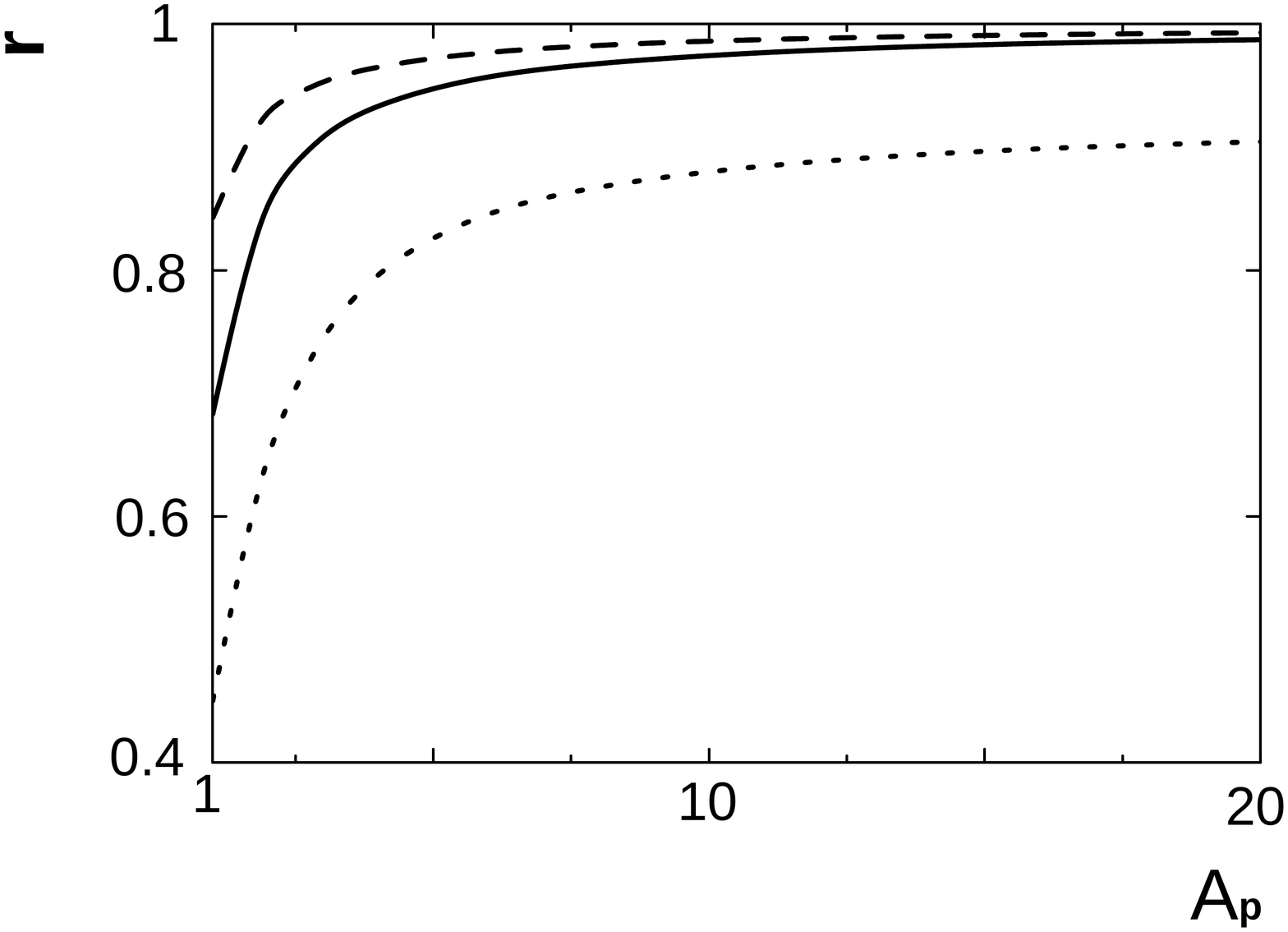}
\caption{ The CE strangeness to entropy ratio divided by 
the corresponding ratio in the GCE
is shown as a function of $A_p$.
The double ratio (Eq.~(\ref{r})) is calculated at the beginning of the mixed phase
$\sqrt{s_{NN}} \approx 7.3$~GeV
(solid line), below the mixed phase region $\sqrt{s_{NN}}=5$~GeV (dotted line),
and above the mixed phase region $\sqrt{s_{NN}}=20$~GeV (dashed line).
  \label{fig-r}
}
\end{figure}

In Fig.~\ref{fig-r} the ratio
\eq{\label{r}
r~\equiv~ \frac{[n^s/s]_{\rm CE}}{[n^s/s]_{\rm GCE}}
}
is shown as a function of $A_p$  
at three collision energies in the vicinity of the transition region.
It is seen that the CE suppression of the strangeness
to entropy ratio depends strongly on the number of  participants $2A_p$.
With increasing
$A_p$ the CE suppression decreases. At $\sqrt{s_{NN}}>10$~GeV the suppression
parameter (\ref{r}) is close to unity already for $A_p>10$.
The CE suppression increases with decreasing collision energy, when the total
number of strange particles is small. This is indicated by the dotted
line in Fig.~\ref{fig-r} which is calculated for collisions at $\sqrt{s_{NN}}=5$~GeV.

\begin{figure}[!htb]
\includegraphics[width=0.49\textwidth]{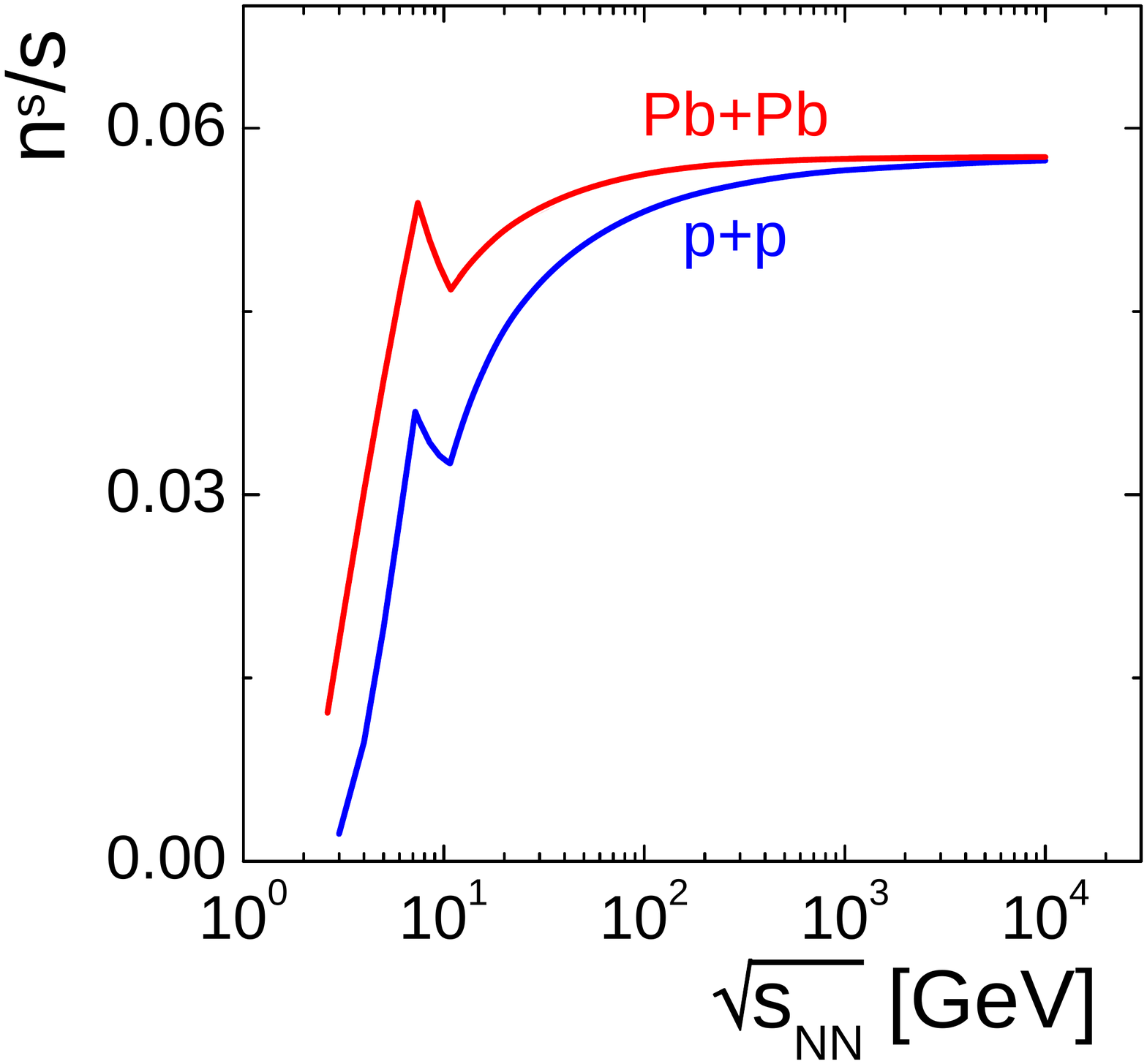}
\includegraphics[width=0.49\textwidth]{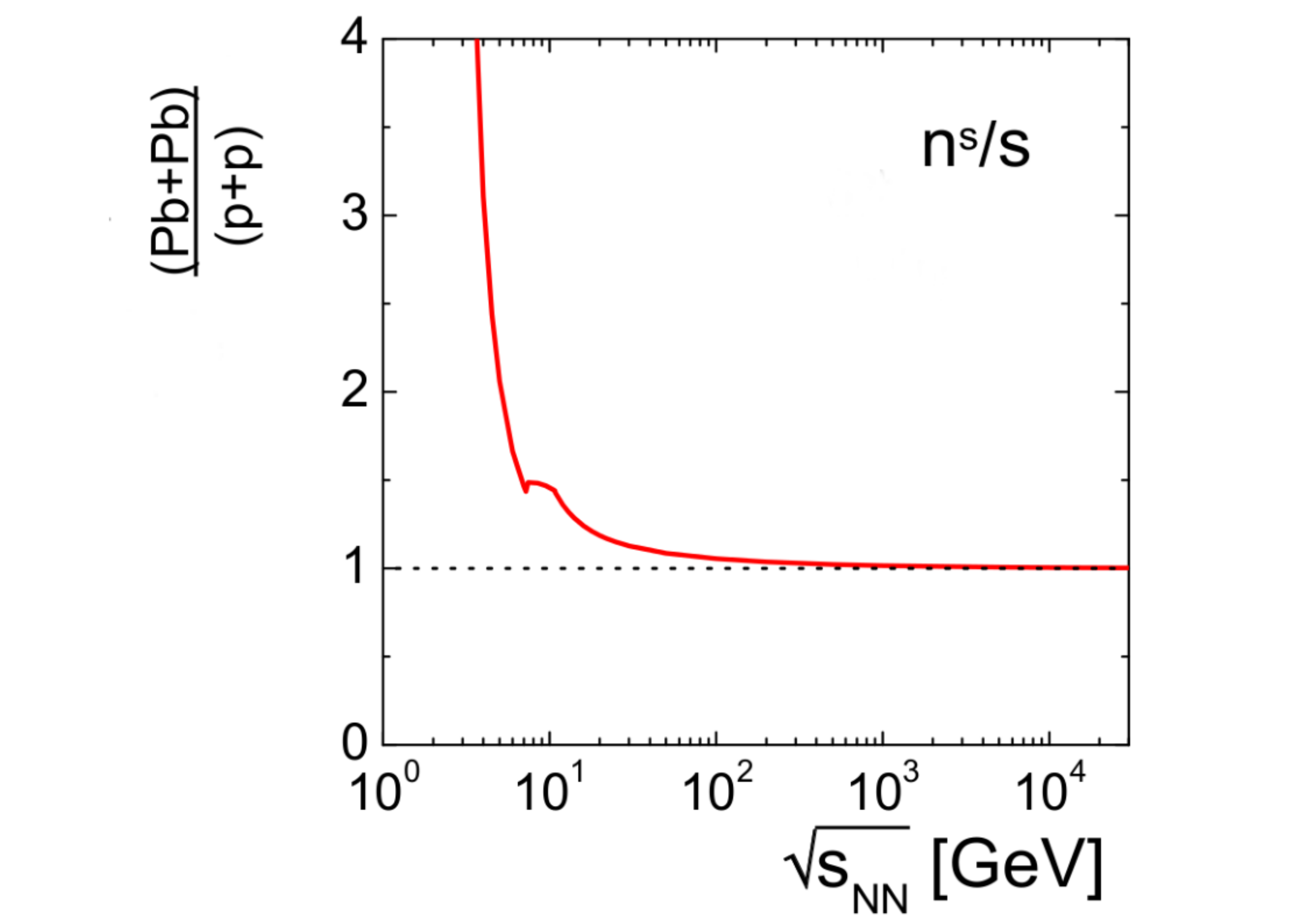}
\caption{ $Left$: 
Collision energy dependence of the strangeness to entropy ratio
calculated within the SMES for central Pb+Pb collisions (the GCE result)
and inelastic p+p interactions (the CE result for $A_p = 1$).
The ratio is plotted up to the LHC energies.
$Right$:
The ratio of strangeness to entropy ratios calculated for central
Pb+Pb collisions and inelastic p+p interactions with the SMES
as a function of collision energy.  
\label{fig-horn-lhc}
}
\end{figure}

Finally, the strangeness to entropy ratio calculated for central Pb+Pb
collisions (the GCE result) and inelastic p+p interactions
(the CE with $A_p = 1$ result)
is plotted in Fig.~\ref{fig-horn-lhc} as a function of collision energy
up to the LHC energies. The left plot shows the two ratios separately, whereas
the right one presents their ratio.
The energy dependence predicted by the SMES is only qualitatively similar
to the measured one (Fig.~\ref{fig:onset}).
Clearly the SMES, the  simplest model of the onset of deconfinement, has to be
significantly modified in order to reach a quantitative agreement
with the data.

\section{Summary}
\label{sec-sum}

This paper introduces the exact
strangeness conservation in  the Statistical Model of the Early Stage~\cite{GG} of
nucleus-nucleus collisions.  This allows
to calculate the energy dependence of the strangeness to entropy ratio
for collisions of protons and small nuclei at the CERN SPS energies.
The extension of the model is motivated by the
recent results of the NA61/SHINE experiment at the CERN SPS on hadron production in inelastic p+p
interactions~\cite{NA61-s}, which suggest that the deconfinement may take place also in
this reaction.

The CE treatment of the strangeness production leads to the well known effect -
the total number of strange and anti-strange particles is reduced in comparison
to that obtained within the GCE at the same values of volume and energy density.
However, the calculations show only small modifications of the system temperature, pressure,
and entropy density.
Thus, the strangeness to entropy
ratio is significantly reduced in small systems.
The smaller  the collision energy, the smaller is the
total number of strange particles, and, thus, the stronger
is the CE strangeness suppression.
In the region of the mixed phase,
$\sqrt{s_{NN}}=7-11$~GeV, the strangeness to entropy ratio in p+p interactions
is found to be approximately two times smaller than  in central Pb+Pb collisions.
Note that the CE suppression  becomes quite small already for central collisions of
intermediate size nuclei and it is negligible for central Pb+Pb collisions.
The calculated collision energy dependence of the strangeness to entropy ratio
in p+p interactions
is qualitatively similar to the one measured by the NA61/SHINE collaboration~\cite{NA61, NA61-s}
for the $K^+$ to $\pi^+$ ratio (see Fig.~\ref{fig:onset}).
However,
a quantitative comparison between the model and the data requires
further modifications of the model and thus being beyond the
scope of this paper.

\begin{acknowledgments}
We would like to thank Marysia Gazdzicka for
corrections to the paper.
This work was supported by
the National Science Centre of Poland (grant
UMO-2012/04/M/ST2/00816),
the German Research Foundation (grant GA 1480\slash 2-2) and
and the National Academy of Sciences of Ukraine, Research
Grant ZO-2-1/2015.
\end{acknowledgments}


\begin{thebibliography}{105}

\bibitem{Fl}
 W.~Florkowski,
  {\it Phenomenology of Ultra-Relativistic Heavy--Ion Collisions}
  World Scientific,
  ISBN: 9814280666,
  436 pages,
  2010.
\bibitem{Afanasev:2000dv} 
  S.~V.~Afanasev {\it et al.}  [NA49 Collaboration],
  CERN-SPSC-2000-035, CERN-SPSLC-P-264-ADD-7.


\bibitem{GG} M. Gazdzicki and M.~I.~Gorenstein, Acta Phys. Pol. {\bf B30}, 2705 (1999).

\bibitem{:2007fe}
  C.~Alt {\it et al.}  [NA49 Collaboration],
  Phys.\ Rev.\  C {\bf 77}, 024903 (2008).


 \bibitem{:2007fe1}
 S.~V.~Afanasiev {\it et al.}  [NA49 Collaboration],
  Phys.\ Rev.\ C {\bf 66}, 054902 (2002).


\bibitem{rustamov} A. Rustamov, Central Eur. J. Phys. {\bf 10}, 1267 (2012).


\bibitem{review}
  M.~Gazdzicki, M.~I.~Gorenstein, and P.~Seyboth,
  Acta Phys.\ Polon.\ B {\bf 42}, 307 (2011); \\
  M.~Gazdzicki, M.~I.~Gorenstein, and P.~Seyboth,
  Int. J. Mod. Phys. E {\bf 23}, 1430008 (2014).

\bibitem{bag} J. Baacke, Acta Phys. Polon. B
{\bf 8}, 625 (1977); \\
 E. V. Shuryak, Phys. Rept. {\bf 61}, 71 (1980); \\
J. Cleymans, R. V. Gavai and E. Suhonen, Phys. Rept. {\bf 130}, 217 (1986).

\bibitem{NA61} N. Abgrall {\it et al.} [NA61 Collaboration], Eur. Phys. J. C 74, 2794 (2014).

\bibitem{NA61-s} N. Abgrall {\it et al.},
CERN-SPSC-2014-031, SPSC-SR-145.

\bibitem{landau}  L. D. Landau, Izv. Akad. Nauk Ser. Fiz. {\bf 17}, 51 (1953);
S.~Z.~Belenkij and L.~D.~Landau, Nuovo Cim. Suppl. {\bf 3S10}, 15 (1956)
[Usp. Fiz. Nauk {\bf 56}, 309 (1955)].


\bibitem{eta} L. P. Csernai, J. I. Kapusta, and L.~D.~McLerran, Phys. Rev. Lett.
{\bf 97}, 152303 (2006).

\bibitem{eta1}
M.~I.~Gorenstein, M.~Hauer,  and O.~N.~Moroz, Phys. Rev. C {\bf 77}, 024911 (2008).

\bibitem{CE}
J. Rafelski and M. Danos, Phys. Lett. B {\bf 97}, 279 (1980).

\bibitem{CE1} J. Clymans, K. Redlich, and E. Suhonen,
Z. Phys. C {\bf 51}, 137 (1991).

\bibitem{CE2} F. Becattini, Z. Phys. C {\bf 69}, 485 (1996) and Nucl. Phys. Proc.
Suppl. {\bf 92}, 137 (2001); \\
F. Becattini and U. Heinz, Z. Phys. C {\bf 76}, 269 (1997).

\bibitem{CE3} M.I. Gorenstein,
M. Ga\'zdzicki, and W. Greiner, Phys. Lett. B {\bf 483}, 60 (2000); \\
M.I. Gorenstein, A.~P.~Kostyuk,
H.~St\"ocker, and W. Greiner, Phys. Lett. B {\bf 509}, 277 (2001); \\
M.I. Gorenstein, W. Greiner,
and A.~Rustamov, Phys. Lett. B {\bf 731}, 302 (2014).


\end{thebibliography}
\end{document}